\documentclass{aa}
\usepackage{txfonts}
\usepackage{psfig,latexsym,amssymb,times}

\usepackage[english]{babel}
\begin{document}
\title{{\it Beppo}SAX and multiwavelength observations of BL Lacertae in 2000}
\author{M. Ravasio\inst{1}
\and G. Tagliaferri\inst{1}
\and G. Ghisellini\inst{1}
\and F. Tavecchio\inst{1}
\and M. B\"ottcher\inst{2}
\and M. Sikora\inst{3}
}
\offprints{M.Ravasio (ravasio@merate.mi.astro.it)}
\institute{Osservatorio Astronomico di Brera, Via Bianchi 46, I-23807
Merate, Italy 
\and Department of Physics and Astronomy, Clippinger 339, Ohio University, Athens, OH 45701
\and Nicolaus Copernicus Astronomical Center, Bartycka 18,00--716 Warsaw, Poland
}
\date{Received ...; accepted ...}
\titlerunning{{\it Beppo}SAX and multiwavelength observations of BL Lacertae in 2000}
\authorrunning{Ravasio et al.\ }
\abstract{ 
We present two {\it Beppo}SAX observations of BL Lac (2200+420) 
as part of a multiwavelength campaign performed in 2000. 
The source was in different states of activity: in June,
 the X--ray spectrum was faint and hard ($\alpha \sim 0.8$),
 with positive residuals towards low energies.
In October  we detected the highest [2--10] keV 
flux ever measured for the source. During this observation, 
the spectrum was soft ($\alpha \sim 1.56$)
up to 10 keV, while above this energy a hard component was dominating. 
The {\it Beppo}SAX data are confirmed by simultaneous RXTE short observations.
During the first observation the soft X--ray flux was variable on 
timescales of a few hours, while the hard X--ray flux was almost constant.
During the second observation, instead, the soft spectrum 
 displayed an erratic behaviour with large variations (up to factors 3--4)
 on timescales smaller than 2 hrs. The analysis of the multiwavelength
SED of October evidenced an intriguing feature: the optical and X--ray sections
of the SED are misaligned, while in the prevailing standard picture,
they are both thought to be produced via synchrotron emission.
We suggested four scenarios to account
for this discrepancy:
 a higher than galactic dust--to--gas ratio towards the source,
the first detection of bulk Compton emission in the X--ray band,
the presence of two synchrotron emitting regions located at different 
distances from the nucleus,
the detection of a Klein--Nishina effect on the synchrotron spectrum.
We evidenced the  favorable and critical points of each scenario, but,
 at present, we cannot discriminate between them.
}
\maketitle
\section{Introduction}
Blazars are radio--loud  Active Galactic Nuclei producing variable non--thermal
radiation  in relativistic jets oriented close to the line of sight:
the emission is therefore beamed and Doppler boosted (Blandford \& Rees, 1978).
They are characterized by a Spectral Energy Distribution (SED)  
displaying two broad  features: 
the first, extending  from radio to UV/X--ray is usually ascribed to
synchrotron emission;
the second, ranging from X--ray to $\gamma$--ray, 
sometimes up to TeV energies, is attributed to inverse Compton scattering 
of seed photons by the same population of synchrotron emitting electrons.
The seed radiation field could be constituted by the synchrotron photons
 themselves (SSC model, Maraschi, Ghisellini \& Celotti, 1992) 
or by external photons, produced by the accretion
disk (Dermer \& Schlickeiser, 1993), by the Broad Line Region 
(Sikora, Begelmann \& Rees, 1994; Ghisellini \& Madau, 1996) or by hot dust 
(Blazejowski et al., 2000; Arbeiter et al., 2002).
 Different contributions of these fields can 
explain the observed blazar spectra.
The BL Lac subclass  is characterized by the 
absence or weakness of broad emission features
 and is divided in HBL and LBL (High and Low energy peaked BL Lacs)
according to
 the radio to X--ray flux ratio (Padovani \& Giommi, 1995).\\
 BL Lac itself (1ES 2200+420)  has been classified as a LBL on the basis of 
 its radio--to--X broad band spectral index $\alpha_{rx} = 0.85$
 (Sambruna et al., 1996).
It was first identified as the optical counterpart of the radio source 
VRO 42.22.01 by Schmitt (1968); the presence of weak narrow emission lines 
in its spectrum allowed an accurate measurement of the redshift $z=0.069$
 of the host elliptical galaxy (Miller \& Hawley, 1977; Miller et al., 1978).
 It shows superluminal motions on m.a.s. scale
 ($\beta_{app}\sim 3-4 ~h^{-1}$, Mutel et al., 1990; 
$\beta_{app} \sim 2.2-5.0 ~h^{-1}$, Denn et al., 2000). 
In spite of the  definition of BL Lac objects as having 
featureless continua, in 1995, a survey 
performed by Vermeulen et al. (1995) revealed the presence of 
an H$\alpha$ emission line with equivalent width of 7 \AA,
 confirmed by subsequent observations (Corbett et al., 1996).
 During the same year  EGRET, aboard the Compton Gamma Ray Observatory,
 detected a 4.4 $\sigma$ excess above 100 MeV from its direction  
(Catanese et al., 1997). 
In the summer of 1997, BL Lac entered an exceptional flaring state with 
the highest X--ray flux ever recorded 
(Sambruna et al., 1999; Madejski et al., 1999)
 and a $\gamma$--ray flux 4 times higher than in 1995 (Bloom et al., 1998) .\\
In the X-ray band, BL Lac has been detected for the first time in 1980 
by the IPC (0.1-4 keV) and the MPC (2-10 keV) aboard the Einstein Observatory 
(Bregman et al., 1990). Since then the source has been observed many times
by different satellites such as EXOSAT (Bregman et al., 1990), GINGA 
(Kawai et al., 1991), ROSAT (Urry et al., 1996; Madejski et al., 1999), 
ASCA (Sambruna et al., 1999; Madejski et al., 1999), 
RXTE (Madejski et al., 1999) 
and finally {\it BeppoSAX}, which observed it in 1997 (Padovani et al., 2001) 
and twice in 1999 (Ravasio et al., 2002).\\ 
During the second half of 2000, from July to December,
the source has been the object of an
intensive multiwavelength campaign (B\"ottcher et al., in prep.) which 
included two  X-ray observations performed by {\it Beppo}SAX and
was supplemented by  a continuous long--term
monitoring program by the Rossi X--ray Timing Explorer (RXTE), with 3 short 
pointings per week (Marscher et al., in prep.).\\ 
During this campaign  BL Lac has been observed  in the radio band
by the telescopes of the University of Michigan and of the
Mets\"ahovi Radio Observatory,
 while in the optical band it has been observed
almost continuously by  24 telescopes in the context of an
extensive WEBT campaign (Villata et al., 2002).
Finally, HEGRA set an upper limit of 25\%  of the Crab flux
above 0.7 TeV, after having accumulated a total of 10.5 h of
 on-source time during the autumn of 2000 (Mang et al., 2001).
In this paper we will analyze in detail the {\it Beppo}SAX data
of this campaign,  comparing them 
with RXTE simultaneous ones and
discussing the spectral and temporal behaviour of BL Lac in the X--ray band
and in the whole radio--to--TeV energy range.

\section{{\it Beppo}SAX observations}

Thanks to its uniquely wide energy range (0.1-200 keV), 
the italian--dutch satellite {\it Beppo}SAX represents an ideal
 experiment for looking at
blazars and expecially objects such as BL Lac, since it can
 detect the transition between the synchrotron and inverse Compton 
 components of the SED 
(Tagliaferri et al., 2000; Ravasio et al., 2002).
Therefore it allows to compare the simultaneous behaviour 
of extremely different parts of the emitting electron distribution.
Boella et al. (1997 and references therein) report an extensive  summary 
of the  mission.\\
{\it Beppo}SAX observed BL Lac (1ES 2200+420) twice during  2000, since the 
 July 26--27 measurements were soon interrupted 
(on--source time $=1.69 \times 10^4$ s). Therefore we were given a second 
chance and a new observation started in October 31 lasting until November 2,
 with a duration of $2.49 \times 10^4$ s.
In Table \ref{tab1} we report  the exposures 
and the mean count rates for each {\it BeppoSAX} instrument.\\ 
\begin{table*}[t]
\begin{center}
\begin{tabular}{ccccccc}
\hline
\hline
 & \multicolumn{2}{c}{\bf{LECS}}  & \multicolumn{2}{c}{\bf{MECS}} & \multicolumn{2}{c}{\bf{PDS}}  \\
\hline
\hline
Date & exposure & count rate$^a$ & 
 exposure  &  count rate$^b$ & 
 exposure &  count rate$^c$ \\
 &  (s) &   $\times 10^{-2}$ cts s$^{-1}$ &
 (s) &  $\times 10^{-2}$ cts s$^{-1}$ & 
 (s) & $\times 10^{-2}$ cts s$^{-1}$ \\
\hline
26-27 July 2000 & 16917 & $4.99 \pm 0.20$ & 23309 & $8.23\pm0.21$  & $1.04\times10^4$ & $-1.20\pm5.73$ \\
\hline
31 October- & & & & & & \\
2 November 2000 & 24899 & $28.5\pm0.38$ & 33661 & $33.5\pm0.33$ & $1.88\times10^4$ & $12.92\pm4.27$\\
\hline
\end{tabular}
\caption{Journal of {\it BeppoSAX} observations. $^a$: 0.1-10 keV; $^b$: 1.5-10 keV;
$^c$: 12-100 keV.}
\label{tab1}
\end{center}
\end{table*}
%
We performed our analysis on linearized and cleaned event files available 
at the {\it Beppo}SAX Science Data Center (SDC) online archive
 (Giommi \& Fiore, 1998) 
using the software contained  in the FTOOLS Package
(XIMAGE 2.63c, XSELECT 1.4b, XSPEC 10.00) and XRONOS 4.02. Data from 
MECS2 \& MECS3 were merged by the SDC team in a single event file.
 Using XSELECT we extracted spectra and light curves from circular regions
around the source of 8 and 4 arcmin radius for LECS and MECS, respectively. 
We extracted event files also from off--source circular 
regions, in order to monitor 
the background behaviour during our measurements. Since the LECS and MECS 
backgrounds are not uniformly distributed across the detectors, after 
having checked the constancy of the extracted background light curves, 
we choose to use the background files obtained from long blank field exposures,
 available at the SDC public ftp site 
(Fiore et al., 1999; Parmar et al., 1999).  
\subsection{Spectral analysis}
The spectral analysis was performed with XSPEC 10.0, 
using the updated response matrices and blank-sky background files (01/2000)
available at the SDC public ftp site.
The LECS/MECS normalization factor was fixed at 0.72, 
while the PDS/MECS was fixed at 0.9 to be consistent
 with our previous works (Ravasio et al., 2002):
 these values are within the acceptable range indicated by SDC 
(LECS: 0.67-1, PDS: 0.77-0.93, Fiore et al., 1999).\\

\subsubsection{July 26-27}

\begin{figure}
\begin{center}
\psfig{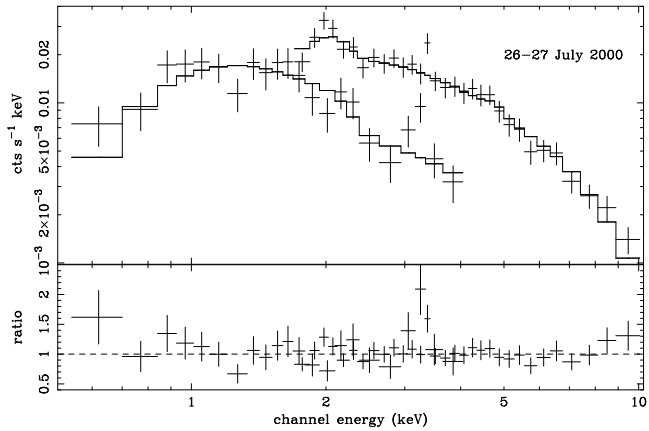}
\caption{LECS+MECS 26-27 July spectrum of BL Lac.
The spectrum is modeled with a power law and a fixed 
 N$_{\rm H}= 2.5\times 10^{21}$ cm$^{-2}$. 
We choose this value to be consistent with previous work
 (Ravasio et al., 2002; Madejski et al., 1999; Sambruna et al., 1999)}
\label{pow-j2000}
\end{center}
\end{figure}

During this observation, the source was not detected by PDS 
because of the short on--source time and because of the intrinsic weakness 
of the source itself.
For the same reasons and because of the high galactic absorption, 
the detection was uncertain also at LECS low energies: 
therefore we proceeded to the analysis only in the [0.6-10] keV range,
fitting the extracted spectrum  with a single power law model. \\
We repeated the procedure three times,  letting the interstellar absorption
 parameter either free to vary,  or fixed  at the value
 N$_{\rm H}=3.6\times10^{21}$ cm$^{-2}$.
The latter is obtained adding  
the galactic value from 21 cm measurements
N$_{\rm H} = (2.0 \pm 1.0) \times 10^{21}$ cm$^{-2}$ (Elvis et al., 1989) 
and  the absorption due to the
molecular clouds observed along the line of sight (Lucas \& Liszt, 1993).
Finally, we used the fixed  value 
N$_{\rm H}=2.5\times10^{21}$ cm$^{-2}$, to be consistent with previous works 
(Ravasio et al., 2002; Sambruna et al., 1999; Madejski et al., 1999).\\
Letting the absorption parameter N$_{\rm H}$ free, the 0.6-10 keV spectrum 
is well fitted by a single power law  with spectral index 
$\alpha = 0.75 \pm 0.15$ ($\chi_r^2/d.o.f.=0.96/50$). 
We obtained N$_{\rm H} = 2.0 \pm 1\times 10^{21}$ cm$^{-2}$,
the large uncertainties are due to  the low count rate
 in the soft X-ray band.\\
A single power law model with N$_{\rm H}$ fixed
 at $3.6\times 10^{21}$ cm$^{-2}$ leaves large positive residuals below $\sim 1$ keV. 
This suggests that we are using an exceedingly high absorption column 
or that near the low boundary of our energy range
we are detecting the transition between 
a soft and a hard spectral component.\\
These residuals become less relevant adopting the intermediate value 
N$_{\rm H} = 2.5 \times10^{21}$ cm$^{-2}$ (fig. \ref{pow-j2000}):
in this case a single power law model
 with $\alpha_X=0.8\pm 0.1$ fits the data well
 ($\chi^2_r/d.o.f. = 0.96/51$).\\
According to an F-test, a broken power law model 
does not  improve significantly the quality of the fit.  
 The best fit parameters  for each  model  are shown in table \ref{tab2},
 which reports also the flux at 1 keV 
and the integrated flux in the 2-10 keV energy range. 

\begin{table*}
\begin{center}
\begin{tabular}{lcccclc}
\hline
\hline
\multicolumn{7}{c}{\bf{26-27 July 2000}}\\
\hline
\hline
N$_{\rm H}$ & $\alpha_1$ & E$_b$ & $\alpha_2$  & F$_{1 keV}$ & F$_{2-10 keV}$ & $\chi^2_r/d.o.f.$ \\
$\times 10^{21}$ cm$^{-2}$ & & keV & & $\mu$Jy & erg cm$^{-2}$ s$^{-1}$ & \\
\hline
free & & & & & & \\
$2.0 \pm 1$ & $0.75 \pm 0.15 $ & & & 1.1 & $ 5.8 \times10^{-12}$ & 0.96/50\\
\hline
fixed & & & & & & \\
3.6 & $0.9\pm0.1$ & & & 1.3 & $5.8\times10^{-12}$ & 1.07/51\\
\hline
fixed & & & & & & \\
2.5 & $0.8\pm0.1$ & & & 1.1 &  $5.8 \times10^{-12}$ & 0.96/51\\
\hline
\hline
 & & & & & & \\
\multicolumn{7}{c}{\bf{31 October - 2 November 2000}}\\
\hline
\hline
N$_{\rm H}$ & $\alpha_1$ & E$_b$ & $\alpha_2$ & F$_{1 keV}$ & F$_{2-10 keV}$ & $\chi^2_r/d.o.f.$ \\
$\times 10^{21}$ cm$^{-2}$ & & keV & & $\mu$Jy & erg cm$^{-2}$ s$^{-1}$ & \\
\hline
\multicolumn{7}{c}{Power law models}\\
\hline
free & & & & & & \\
$3.0\pm0.3$ & $1.63 \pm 0.06 $ & & & 13.0 & $2.1\times10^{-11}$ & 0.69/76\\
\hline
fixed & & & & & & \\
3.6 & $ 1.71 \pm 0.03$ & & & 14.6 & $2.1\times10^{-11}$  & 0.96/77\\
\hline 
fixed & & & & & \\
2.5 & $1.56\pm0.03$ & & & 11.9 & $2.1\times10^{-11}$ & 0.87/77 \\
\hline
\multicolumn{7}{c}{Broken power law models}\\
\hline
\hline
fixed & & & & & & \\
2.5 & $1.45^{+0.1}_{-0.5}$ & $2.2^{+0.7}_{-1.3}$ & $1.65\pm0.08$ & 11.4 & $2.1\times10^{-11}$ & 0.68/75 \\
\hline 
\hline
\end{tabular}
\caption{LECS + MECS spectrum best--fit parameters.}
\label{tab2}
\end{center}
\end{table*}

\subsubsection{October 30- November 2}

\begin{figure}
\centerline{
\vbox{
\psfig{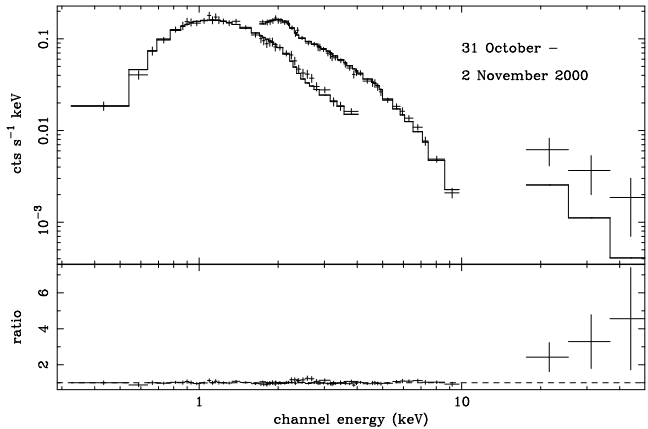}
}}
\caption{LECS+MECS+PDS  31 October--2 November spectrum of BL Lac.
Both a single and a broken power law model cannot reproduce 
the whole observed spectrum. 
PDS data lie above the power--law best fit 
of the of LECS+MECS spectrum.
The LECS+MECS spectrum is well fitted by a convex broken power law.}
\label{pow+pds-oct} 
\end{figure}

The October observation is more interesting because
 of the longer duration and the higher state of the source.
This allows us to analyze a wider spectral range, from
 0.3 keV up to 50 keV, thanks to the detection by the PDS.\\
 As before, we performed the fits  letting N$_{\rm H}$ free, 
then fixing it to its maximum value and finally to 
N$_{\rm H}=2.5\times 10^{21}$ cm$^{-2}$.
In the first case,  the LECS + MECS spectrum
 was well fitted by a soft single power law
 model, with energy index  $\alpha_X = 1.63 \pm 0.06$ 
(N$_{\rm H} = 3.0\pm 0.3 \times 10^{21}$ cm$^{-2}$).
A similar result is obtained also keeping   
N$_{\rm H}$ fixed to $2.5 \times10^{21}$ cm$^{-2}$, while the model
with the highest absorption  value leaves positive residuals increasing
towards low energies.\\
We repeated the procedure using a broken power law model and  
fixing N$_{\rm H} =2.5\times 10^{21}$ cm$^{-2}$, 
for consistence with previous works (e.g. Ravasio et al., 2002).
The LECS+MECS spectrum in the latter case is well fitted 
by a convex curve steepening beyond $\sim 2$ keV.
The best--fit parameters  of each  LECS + MECS spectral model 
are listed in Table \ref{tab2}.
An F-test suggests that the addition of  two parameters 
gives a $ 99.9 \%$ probability of improving the quality of the fit.\\
Subsequently we included also  PDS data in the analysis.
 A  single power law model  leaves large positive residuals 
towards high energies (fig. \ref{pow+pds-oct}): 
the [0.1--50] keV spectrum seems to be concave. PDS data lie largely above
the extrapolation of the LECS and MECS spectrum: this can be explained
as the rising of a second, hard spectral component. 
However, the error bars are too large and we cannot constrain
a second component in the model.
Thus, in order to have an idea of this hard 
component spectral shape, we fitted the MECS + PDS spectra with a broken 
power law model, keeping the low energy spectral index fixed to the 
value obtained from the fit of the LECS + MECS spectrum ($\sim 1.65$, 
independently from the absorption parameter chosen).
The best fit of the two extra parameters are E$_b \sim 9.9$ keV 
and $\alpha_{\rm {PDS}} \sim 0.56$,
but we are not able to estimate the uncertainties. 

\subsubsection{Simultaneous RXTE observations}                              
                           
In addition to {\it Beppo}SAX,  the 2000 campaign 
was covered in the X--ray band also by  Rossi X--ray Timing Explorer
(Bradt, Rotschild \& Swank, 1993), which provided three
short exposures a week, covering the whole 
duration (Marscher et al, in prep.). 
We analyzed  the two RXTE pointings temporally closest
 to the {\it Beppo}SAX ones. In July the two observations were 
exactly simultaneous while in November RXTE was lagging 
{\it Beppo}SAX only by 1 hour.
 Therefore we are  given the opportunity to test our 
results with a totally independent set of data.\\
We restrict our analysis to PCA data (Jahoda et al., 1996),
an instrument composed by 5 passively collimated independent X--ray 
detectors sensitive to the 2-60 keV range.\\
We choose to compare the  [3-15] keV RXTE spectrum to the [1-10] keV MECS
using a power law model with  fixed absorption parameter  
(N$_{\rm H} = 2.5 \times10^{21}$ cm$^{-2}$).
The model parameters are reported in Table \ref{xte-mecs}, together
with the log of the observations.
\\  
Both RXTE spectra are well fitted by power law models: during the
observation of July, RXTE detected a hard spectrum 
($\alpha= 0.9$), while in October, it detected a soft
component ($\alpha=1.45$). The slope and the normalization 
of the best-fit models of RXTE spectra are consistent with those of the MECS, 
confirming the results obtained from {\it Beppo}SAX spectral analysis.
\begin{table*}
\begin{center}
\begin{tabular}{cccccccc}
\hline
\multicolumn{8}{c}{\bf 26-27 July}\\
\hline
Instrument & Time start & Time end & Duration  &$\alpha$ & F$_{1 keV}$ & F$^{*}_{2-10 keV}$ & $\chi^2_r$/d.o.f.\\
          &            &          &   (sec)   & & ($\mu$Jy)  & ($\times10^{-12}$) & \\
\hline
PCA &  26/7/00  & 26/7/00 & 2208  & $0.9^{+0.7}_{-0.6}$& 1.4 & 6.3 & 0.42/25\\
3-15 keV    & 18:23:44  &  19:00:32 &  & &  & & \\
\hline
MECS & 26/7/2000 & 27/7/2000 & 23309  & $0.8\pm0.1$ & 1.18 & 5.8 & 0.86/43\\
1-10 keV     & 10:12:39  & 06:43:33 &   &  & &  &\\
\hline
\multicolumn{8}{c}{\bf 30 October-2 November}\\
\hline
PCA & 2/11/2000 & 2/11/2000 & 2000  & $1.45^{+0.3}_{-0.25}$ & 10.3 & 19.7 & 0.45/25 \\
3-15 keV    & 10:56:16  & 11:29:36  &         &       &      & &  \\
\hline
MECS & 31/10/2000 & 2/11/2000 & 33661  & $1.6\pm0.05$ & 12.7 & 19.7 & 0.61/58 \\
1-10 keV     &  20:46:55 & 09:59:28 & & & & & \\
\hline
\end{tabular}
\caption{$^{*}$: erg cm$^{-2}$ s$^{-1}$. 
RXTE and {\it Beppo}SAX simultaneous observations log and 
spectral fit parameters. We adopted a power law model with
fixed absorption parameter: N$_{\rm H}=2.5\times10^{21}$ cm$^{-2}$.}
\label{xte-mecs}
\end{center}
\end{table*}
\subsection{Temporal analysis}
We performed this analysis with the XRONOS 5.16 package, 
grouping data in different time bins  to reduce uncertainties.
During the observation of July 2000 the source was in a faint state 
of activity so we choose bins of 3600 s.  During the second observation,
 the source emission was  stronger so we could rebin our data in smaller 
 bins of 600 s. In our analysis we excluded 
the bins with less than $20\%$ of effective exposure time, to 
reduce weighting errors.

\subsubsection{26-27 July}
\begin{figure}
\psfig{figure=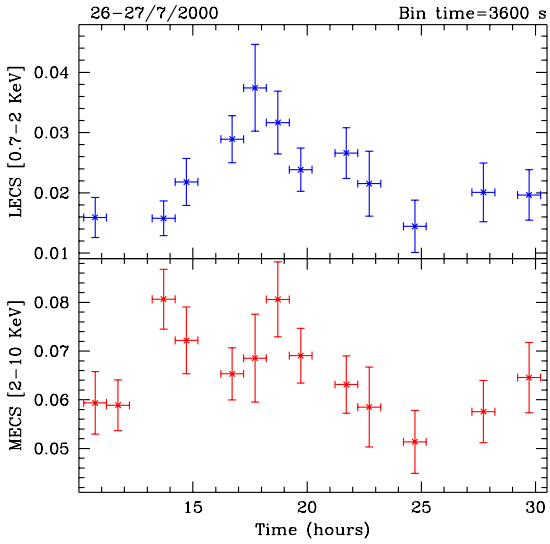,width=8.5cm}
\caption{LECS 0.7-2 keV (top panel), MECS 2-10 keV (bottom panel) light curves 
of BL Lac during the July 2000 {\it Beppo}SAX observation. Note that in the LECS
range the source displays larger variability than at higher energies.}
\label{2curve-JULY}
\end{figure}

In order to have information on the behaviour of the source 
at different X-ray wavelenghts we extracted a soft ([0.7--2] keV, LECS) 
and a hard ([2--10] keV, MECS) X--ray light curve.
The source displays large  variability 
in the low energy band: the 0.7-2 keV flux increases by a factor $>2$ 
 on time scales of $\sim 4$ hours, then fades to previous values 
in $\sim 5-6$ hours (Fig. \ref{2curve-JULY}).     
In the harder energy band, variability events are less evident but the source 
is still active, with 
flux variations of $25-30\%$ ($\chi^2$-test constancy probability
 $\sim 10\%$). The two light curves are only marginally correlated:
the null correlation probability is quite high ($\sim 25\%$).\\
This behaviour was noticed also in the June 1999 {\it BeppoSAX} 
observation of BL Lac (Ravasio et al., 2002):
 in the high energy MECS band (4-10 keV)
 the source was not variable, while the [0.3-2] keV LECS
and the [2-4] keV MECS light curves were displaying a  flare
of a factor $\sim 4$. 
\subsubsection{31 October-2 November}
\begin{figure}
\psfig{figure=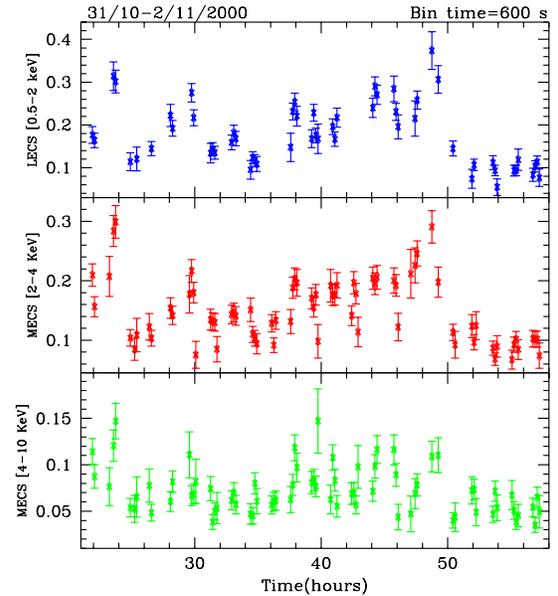,width=8.5cm}
\caption{LECS 0.5-2 keV (top panel), MECS 2-4 keV (mid panel) and MECS 4-10 
keV (bottom panel) light curves
of BL Lac during the October-November 2000 {\it Beppo}SAX observation.}
\label{2curveratio-oct}
\end{figure}

During the October {\it Beppo}SAX observation, BL Lac displayed
very fast and remarkable flux variations in the whole energy range covered 
by the LECS and the MECS (Fig. \ref{2curveratio-oct}). 
The light curves are characterized by a similar, erratic behaviour, with flux 
variations of factors $\sim 3-4$, even on timescales of $\sim 1$ hour.
The three light curves are highly correlated (P$>99.9 \%$) 
even with a short temporal binning of 10 min.\\
We analyzed also the PDS light curve, which turned out to be constant:
 a  test  gives a $\sim 96\% $ constancy probability. 
Because of the large errorbars, however, we would not be able to detect 
variations smaller than a factor of 3.
For BL Lac, this  X--ray behaviour is not unprecedented: 
as already mentioned, during  the June 1999 
{\it Beppo}SAX observation, the  [0.3--2] keV flux  doubled in 20 min
and faded to previous values in a similar time.
 In less than 2 hours a complete flare with a flux variation
of a factor $\sim 4$ was observed. The amplitude of this 
event was highly frequency--dependent: while the flare was extremely 
prominent in the [0.3--2] keV and in the [2--4] keV curves, 
at higher frequencies the flux remained constant.\\
%
The best way to characterize the temporal behaviour of the source would be
the calculation of the power density spectrum: however this tool is not 
appropriate for unevenly sampled light curves like  {\it Beppo}SAX ones. 
The large observational gaps, the short duration of the run
and the limited statistics caused by the faintness of the source
also makes the use of alternative techniques very difficult,
such as the discrete correlation function  (Edelson \& Krolik, 1988) or the 
structure function calculation (Simonetti, Cordes \& Heeschen, 1985).\\
We can still characterize the variability degree of the source 
using two common estimators:
 the {\it normalized excess variance} parameter $\sigma^2_{\rm rms}$
and the {\it minimum doubling timescale} T$_{short}$ (Zhang et al., 1999;
Fossati et al., 2000). The first parameter is defined as the normalized 
difference between the variance of the light curve  and the variance due to
measurement errors: it quantifies the mean variability of the source.
The second parameter, instead, represent a measure of the fastest 
significant timescale of the source (Edelson, 1992).
Assuming that each point of the light curve is described as ($t_i$,$f_i$)
 we define the ``doubling time'' as
\begin{equation}
T_{ij}=\Big|\frac{f_{ij}\Delta T_{ij}}{\Delta f_{ij}}\Big|
\end{equation}
where $\Delta T_{ij}=t_i-t_j$, $\Delta f_{ij}=f_i-f_j$ 
and $f_{ij}=(f_i+f_j)/2$.\\ 
 For each of the discussed energy ranges ([0.5--2] keV--LECS; [2--4] keV--MECS;
[4--10] keV--MECS)
 we produced 4 different light curves
 with time bin  of 500, 1000, 1500 and 2000 sec, respectively.
For each  curve we computed the $\sigma^2_{\rm rms}$ and the average
of the 5 smallest $T_{ij}$ (calculated for each data pair (i,j)
 of the light curve)
with fractional error lower than 25\%.
We define $T_{short}$ as their weighted average.
%
%
In Table \ref{var} we summarize our results.\\
The highest frequency light curve is the least variable, 
as can be noticed also by looking at Fig. \ref{2curveratio-oct}: probably 
this can be  explained partially as an effect of the larger measurement error
caused by  poorer statistics and partially as 
a lower variability at higher energies.\\
 Within the relatively large errors, no significant differences are found
between the $T_{short}$ values of the three light curves:
 the source seems to have
a characteristic minimum variability timescale of $\sim 1.5-2$ hrs.
\begin{table*}
\hspace{-4cm}
\begin{center}
\begin{tabular}{lcccc}
\hline
\hline
\multicolumn{5}{c}{\bf{30/10-2/11 2000}}\\
\hline
\hline
Energy band & $<$Count rate$>^a$ & $\sigma^2_{rms}$$^a$ & $< \sigma^2_{rms}>$$^b$ & T$_{short}^b$ \\
(keV) & ($10^{-2}$ cts/s) & ($10^{-2}$) & ($10^{-2}$) & (ks)\\
\hline
LECS & 17.6 & $15.2$ & $12.8$  & 6.3\\
0.5-2 keV & $\pm7.2$ & $\pm3.3$ & $\pm 1.6$ & $\pm 0.6$ \\
\hline
MECS & 14.7 &11.4 & 10.1 & 5.5 \\
2-4 keV & $\pm 5.3$ & $\pm 2.5$ & $\pm1.5$ & $\pm 0.5$ \\
\hline 
MECS & 7.4 & 7.9 & 7.5 & 7.1 \\
4-10 keV & $ \pm2.5$ & $\pm 2.1$ & $\pm 1.1$ & $\pm 1.3$\\
\hline
\end{tabular}
\caption{Variability parameters. $^a)$ From 600 sec. rebinned light curves.
$^b)$ Weighted mean of data taken from 500, 1000, 1500,
2000 sec. rebinned light curves.}
\label{var}
\end{center}
\end{table*}
\section{Spectral Energy Distributions} 
\begin{figure*}
\begin{center}
\psfig{figure=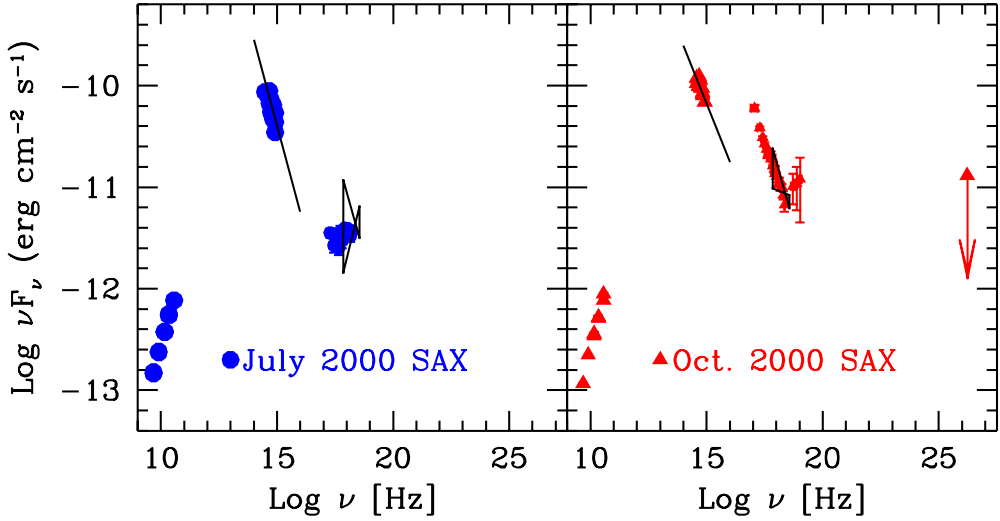,width=14cm}
\caption{Left panel: 26--27 July SED. Right panel: 31 October--2 November
SED. The black butterflies represent RXTE simultaneous data.
The HEGRA upper limit is calculated assuming a spectral 
energy  index $\alpha_{\rm TeV}=1.5$ above 0.7 TeV.
 The optical slopes are calculated on the 
weigthed average UBVRI fluxes (A$_{\rm B}=1.42$).
July fluxes (mJy):  $<{\rm F_I}>=5.17\pm0.07; <{\rm F_R}>=8.12\pm0.01; 
<{\rm F_V}>=12.08\pm0.04; <{\rm F_B}>=16.54\pm0.02; <{\rm F_U}>=22.28\pm0.06$.
October fluxes (mJy): ${<\rm F_I}>=8.26\pm0.09; <{\rm F_R}>=11.96\pm0.05; 
<{\rm F_V}>=17.56\pm0.06; <{\rm F_B}>=22.43\pm0.08; 
<{\rm F_U}>=27.97\pm0.18 $. }
\label{sed-2000}
\end{center}
\end{figure*}

After having performed the spectral analysis we are now able to build 
the SEDs of BL Lac relative to the {\it Beppo}SAX 2000 observations
and to compare them with other historical multiwavelength SEDs.\\
As we have discussed above, during 2000 {\it Beppo}SAX detected the source 
 in two completely different states of activity:
in July the source was detected only in the [0.6--10] keV range, 
displaying a faint (F$_{2-10}=5.8\times10^{-12}$ erg cm$^{-2}$ s$^{-1}$)
hard spectrum. In October--November, instead, BL Lac was displaying 
a very intense (F$_{2-10}=2.1\times10^{-11}$ erg cm$^{-2}$ s$^{-1}$)
soft spectrum up to $\sim 10$ keV, while at higher energies
a hard  component, detected by the PDS up to 45 keV,
 was dominant.\\
Also in the optical band the source was in different states: the increase
of the optical flux, in fact,  was the reason for prolonging the 
multiwavelength campaign beyond August 2000.
 The optical fluxes measured simultaneously to the
 {\it Beppo}SAX observations
differed by 40\% between the two pointings:  in the
core of the campaign (26-27 July) the source average 
R--band magnitude was  m$_R=14.08$, while during the second X--ray run 
 it was m$_R=13.74$ (Villata et al., 2002).
After having dereddened the data using A$_{\rm B}=1.42$ 
(Schlegel et al., 1998),
we calculated the optical spectral indices using weight--averaged 
 UBVRI fluxes (see values in fig. \ref{sed-2000} caption):
when fainter, the spectrum is softer ($\alpha_{opt}\sim1.84\pm 0.01$), while
during the autumn it is harder ($\alpha_{opt}\sim 1.58\pm0.04$).\\
In Fig. \ref{sed-2000} we plot the two simultaneous multiwavelength SEDs
 of BL Lac. 
We report also the upper limit  by HEGRA,
$7.7\times10^{-12}$ photons cm$^{-2}$ s$^{-1}$ above 0.7 TeV 
(Mang et al., 2001), which
is $25 \%$ of the Crab level: this means 
 F$_{0.7 \rm TeV} = 5.13\times10^{-15}\times(\alpha_{\rm TeV})$ Jy, where 
$\alpha_{\rm TeV}$ is the energy index of the TeV spectrum (assumed to be $=1.5$ 
in the plot).
\subsection{Historical SEDs}
As mentioned in the introduction, blazar SEDs are characterized by two 
broad humps extending, the first, from radio to UV/X--rays and the second
from X--rays to $\gamma$-ray or even TeV energies. The first component is
usually interpreted as synchrotron emission from high--energy electrons 
in relativistic motion, while the second is explained as inverse Compton 
scattering of seed photons of still debated origin.\\
\begin{table*}[t]
\begin{center}
\begin{tabular}{|l|c|c|c|c|c|}
\hline
Date &$\alpha_{1keV}$ & F$_{1keV}$ & F$_{2-10 keV}$ & $\alpha_{opt}$ & F$_{\rm V}$ \\
 &  & ($\mu$Jy) & ($10^{-12}$ c.g.s.) & & (mJy)\\
\hline
June 1980 & $2.2\pm1.0$ & 1.6 & 7.1$^{*}$ & $2.2\pm0.2$ & 5.75 \\ 
\hline
December 1983 & (2.2) & 0.8 & & $2.1\pm0.4$ & 9.55 \\
\hline
June 1988 & $0.71\pm0.07$ & 1.27 & 7.7$^{*}$ & $2.2\pm0.1$ & 4.2 \\
\hline 
July 1988 & $1.16\pm0.24$ & 1.35 & 4.1$^{*}$ & $3.21\pm0.02$ & 6.46 \\
\hline
December 1992 & $0.94\pm0.46$ & 0.88 & 3.7$^{*}$ & & \\
\hline
November 1995 & $0.96\pm0.04$ & 2.22 & 8.9 & $0.8\pm0.2$ & 5.7 \\
\hline
July 1997 & $0.59\pm0.03$ & 3.55 & 26 & $1.85\pm0.4$ & 27.9 \\
\hline
November 1997 & $0.89\pm0.13$ & 2.4 & 11.2 & & \\ 
\hline
June 1999 & $1.57\pm0.25$ & 0.77 & 6.5 & $1.2\pm0.2$ & 26.4 \\
\hline
December 1999 & $0.63\pm0.06$ & 0.96 & 12.3 & $1.8\pm0.3$ & 18.6\\
\hline
July 2000 &  $0.8\pm0.1$ & 1.1 & 5.8 & $1.84\pm0.01^a$ & 12.08$^a$ \\
          &  $0.8\pm0.1$ & 1.1 & 5.8 & $0.94\pm0.01^b$ & 25.60$^b$\\ 
\hline
October & $1.45^{+0.1}_{-0.5}$ & 11.4 & 21 & $1.58\pm0.04^a$ & 17.56$^a$ \\
November 2000$$ & $1.45^{+0.1}_{-0.5}$ & 11.4 & 21 & $0.64\pm0.02^b$ & 37.23$^b$  \\
\hline
\end{tabular}
\caption{June 1980 Einstein data: Bregman et al., 1980.
December 1983 EXOSAT data: Bregman et al., 1990.
June 1988--July 1988 GINGA data: Kawai et al., 1991.
December 1992 ROSAT data: Urry et al., 1992.
November 1995 ASCA data: Sambruna et al., 1999. 
July 1997 RXTE data: Madejski et al., 1999. 
November 1997 {\it Beppo}SAX data: Padovani et al., 2001. 
June and December 1999 {\it Beppo}SAX data: Ravasio et al., 2002.
$^*$: from our calculations.
$^a$: using A$_{\rm B}=1.42$. $^b$: using A$_{\rm B}=2.5$: see discussion.} 
\label{history}
\end{center}
\end{table*}
The multiwavelength history of BL Lac follows that of its X--ray
observations which can be traced back to 1980, the year of
 the first X--ray detection, carried out by Einstein (Bregman et al., 1990).
BL Lac was then observed also by EXOSAT (Bregman et al., 1990),
 GINGA (Kawai et al., 1991), ROSAT (Urry et al., 1996; Madejski et al., 1999),
 ASCA (Sambruna et al., 1999; Madejski et al., 1999), 
RXTE (Madejski et al., 1999) and by {\it Beppo}SAX  (Padovani et al., 2001; 
Ravasio et al., 2002).\\
In table \ref{history} we report the published spectral parameters 
for  each X--ray observation, together with simultaneous optical ones.
%
\begin{figure*}
\begin{center}
\hspace{0.5cm}
\psfig{figure=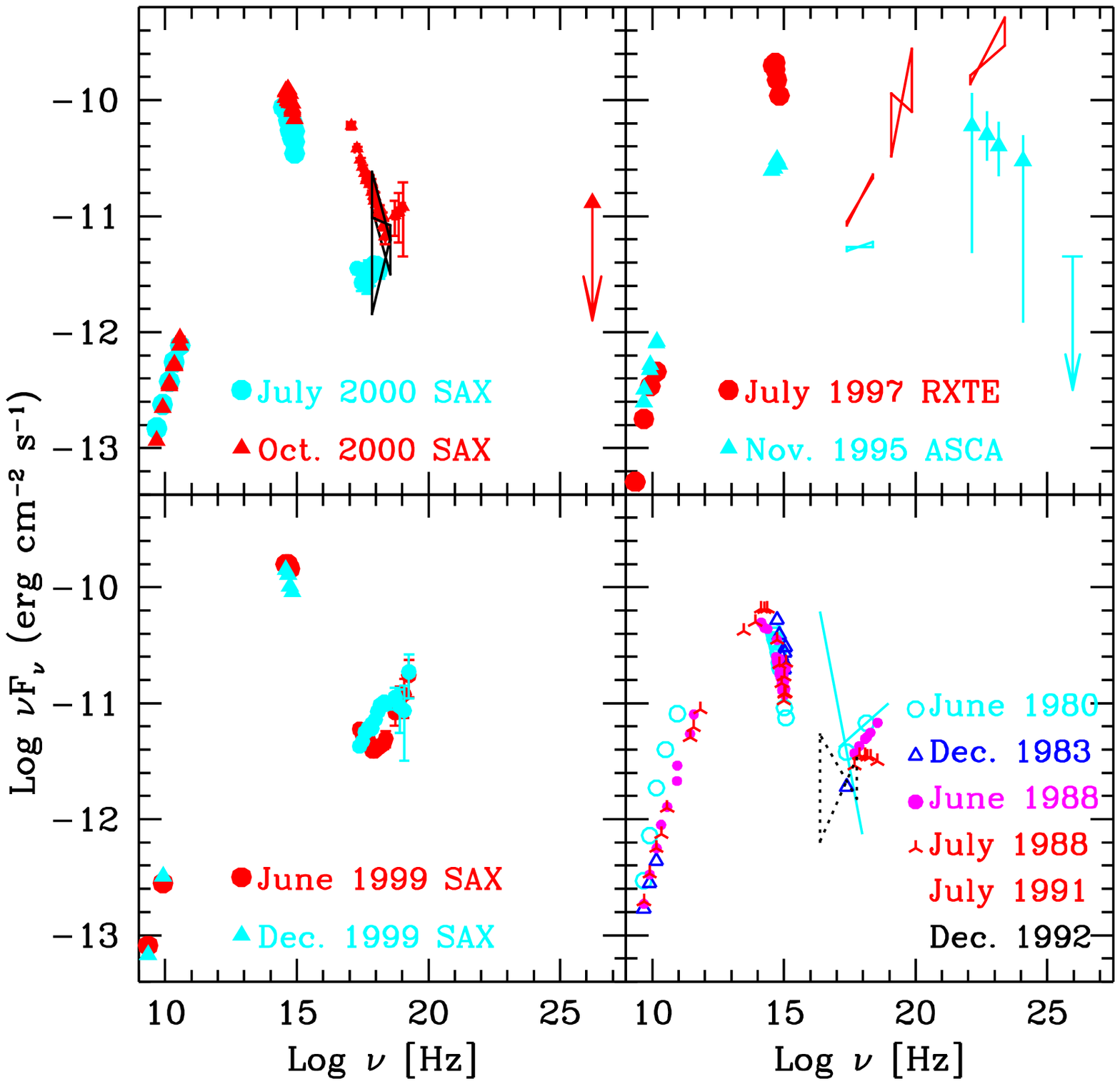,width=14cm}
\caption{Simultaneous multiwavelength SEDs of BL Lac.
Left--top panel: July and October--November 2000 SEDs
 ({\it Beppo}SAX). We have reported optical data corrected for 
 A$_{\rm B}=1.42$, to be consistent with the other 
historical observations.
 Right--top panel: November 1995 (ASCA) and July 1997 (RXTE) SEDs.
Left--bottom panel: June and December 1999 SEDs ({\it Beppo}SAX).
Right--bottom panel: June 1980 (Einstein), December 1983 (EXOSAT),
June and July 1988 (GINGA)  multiwavelength SEDs.
The dotted butterfly represent the December 1982 ROSAT data.}
\label{SED}
\end{center}
\end{figure*}
 The X--ray  spectra of BL Lac exhibit a variety of different shapes:
 sometimes
they are concave, hardening towards high energies, and sometimes 
they are hard in the whole X--ray band.
In 1980, IPC ([0.1-4] keV) and MPC ([2-10] keV) 
aboard Einstein detected quite different spectral indices 
 ($\alpha_1=2.2$; $\alpha_2=0.8$; Bregman et al., 1990) and
a similar shape was detected also by ASCA in November 1995 
(Sambruna et al., 1999) and by {\it Beppo}SAX in June 1999 
(Ravasio et al., 2002), as well as during this campaign.
 In other epochs the X--ray spectrum of BL Lac 
was displaying only a hard component. 
In June and July 1988 GINGA detected  faint X--ray  spectra
which can be attributed to inverse Compton emission
in a leptonic jet model (Kawai et al., 1991),
 similarly to the observation of ROSAT (1982),
to those performed  by {\it Beppo}SAX in November 
1997 (Padovani et al., 2001), December 1999 (Ravasio et al., 2002)
and to our observation of July 2000.
All the  hard X--ray spectra  are similar to each other 
except for  one case: during July 1997 BL Lac was in a high state,
 consistent with  Compton emission,
 with  a 2-10 keV flux $\sim 3-4$ times higher than that observed in
 1995 by ASCA or in June 1999 by {\it Beppo}SAX;
the high state was clearly detected also in 
the $\gamma$--ray by EGRET (Bloom et al., 1997). 
 The  complexity of the behaviour of the source, suggested by 
the data reported in table \ref{history},  is further highlighted
by the comparison of the multiwavelength SEDs:
in fig. \ref{SED} we plot the results of all the multiwavelength
campaigns performed on BL Lac.\\
As evidenced in fig. \ref{sed-2000} \& \ref{SED} 
and in table \ref{history}, 
while during July 2000 we observed the source in a normal state,
 in the end of October BL Lac was very active.
During this run, in fact, {\it Beppo}SAX detected the highest soft X--ray
flux and an integrated [2--10] keV flux which is only slightly smaller
than that of the flare of July 1997, 
when BL Lac was displaying a hard X--ray spectrum. 
\section{Discussion}
X--ray observations are  crucial for blazar study because at these energies 
we can often observe the transition between the  two emission mechanisms.
The two X--ray observations that we presented here clearly
show this:
\begin{itemize}
\item {\bf 26--27 July}: during this observation {\it Beppo}SAX detected
a hard spectrum in the LECS--MECS range, with large positive residuals
below 1 keV. The hard component can be attributed to inverse Compton emission
while the residuals can be explained as the very soft tail of the synchrotron 
component (the data however does not allow us to set limits on the 
spectral index of this component neither to asses its consistency
with the optical data).
 This picture is supported by the temporal behaviour of the source:
 the [2-10] keV light curve represents the inverse Compton spectrum
produced by the less energetic particles and is therefore almost constant
on the timescale of a single observation.
 The [0.7-2] keV curve, instead,  is influenced also by the soft
synchrotron emission produced by highly energetic electrons 
which can account for the observed variability.
\item {\bf 31 October -- 2 November}: also in this observation {\it Beppo}SAX
detected the transition between  two emission mechanisms. The LECS+MECS
spectrum is soft, while the PDS observed a hard component.
In the usual scenario, the hard component is attributed to 
inverse Compton emission, while the soft spectrum is explained as 
synchrotron emission by very energetic electrons
 ($\gamma \sim 2.5 \times 10^4-10^5$ (B/1G)$^{-1/2} (\delta/10)^{-1/2}$)
 which have a very short cooling time
 (t$_{syn} \sim 3\times 10^3-10^4$(B/1G)$^{-3/2} (\delta/10)^{1/2}$ s).
 This could also 
explain the very fast variability detected in the whole LECS--MECS 
energy range.\\
This scenario, however, could be inadequate.
Looking at fig. \ref{sed-2000}, one can notice immediately a strange
feature in the SED of October--November: the X--ray data
lie above the extrapolation of the optical spectrum, while
both of them should be produced by the same emission mechanism, 
the synchrotron.
Even correcting for the  host galaxy contribution (which is almost 
negligible for the  level of activity of BL Lac during our observations)
 we are not able to reconcile X--ray and optical data. 
 However, we can conceive at least four scenarios which under certain 
conditions can explain the observed  spectral ``glitch''. They are:
 I) a variable local absorbing column along the line of sight; 
II)  bulk Compton radiation;
 III) two different synchrotron emitting regions;
 IV) Klein-Nishina effect on the synchrotron spectrum
.\\  
\end{itemize}
\subsection{Differential absorption}
%

\begin{figure}
\begin{center}
\hspace{-1.3cm}
\psfig{figure=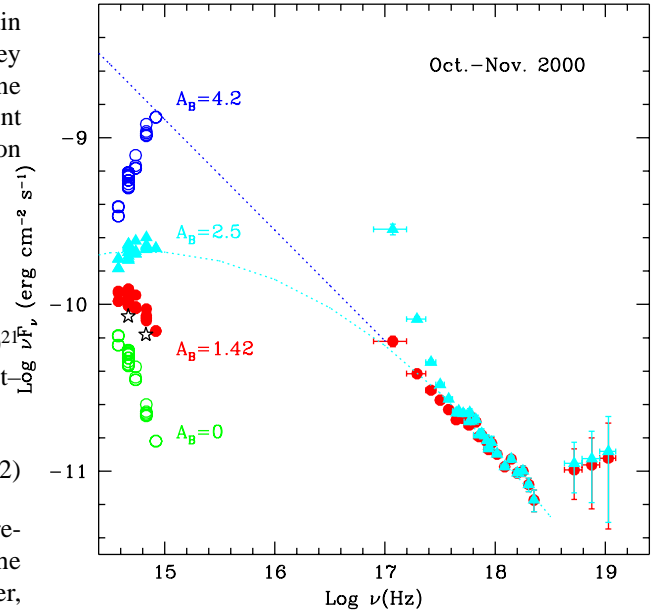,width=10cm}
\caption{Optical to X--ray SED of October--November 2000 {\it Beppo}SAX 
observation. Filled circles represent data corrected for 
A$_{\rm B}=1.42$  corresponding to N$_{\rm H}=2.5\times10^{21}$ cm$^{-2}$. 
Filled triangles represent data corrected for A$_{\rm B}=2.5\equiv \rm N_{\rm H}=4.8\times10^{21}$ cm$^{-2}$. 
Open circles represent raw optical data and optical data dereddened with
A$_{\rm B}=4.2$.
The two stars represent the B and R--band data dereddened with 
A$_{\rm B}=1.42$ and corrected for the host galaxy contribution.
The two dotted lines represent the extrapolation of the {\it Beppo}SAX
spectrum to the optical band and a parabolic curve which 
connects X--ray and optical data  dereddened with A$_{\rm B}=2.5$.
See text for discussion.}
\label{abs}
\end{center}
\end{figure}

From the {\it Beppo}SAX X--ray analysis we had
 N$_{\rm H}=2.5\times 10^{21}$ cm$^{-2}$,
consistent with previous observations: using the dust--to--gas
ratio suggested by Ryter (1996)
\begin{equation}
\rm{A}_V=4.5\times10^{-22}~ \rm{N}_{\rm H} ~\rm{cm}^{-2}
\end{equation}
and using E(B-V)=0.329 as indicated by the NED, 
the corresponding optical absorption is A$_{\rm B}=1.45$, 
 very similar to the value reported by Schlegel et al. (1998),
A$_{\rm B}=1.42$. However, using these values, 
the X--ray spectrum lies largely above the
optical spectrum extrapolation, as illustrated in Fig. \ref{abs}.
Even correcting the optical data 
for the host galaxy contribution the two sections 
do not connect continuously. In Fig. \ref{abs} we plotted
the R and B band BL Lac fluxes corrected using R$_{host}=15.55$ and
B$_{host}=17.15$  (Villata et al., 2002): the misalignement
is even worsened.\\
If we try to account for this discrepancy using different
 absorption values, e.g. A$_{\rm B}=2.5$, and consequently 
N$_{\rm H}=4.8\times 10^{21}$  cm$^{-2}$, 
we still obtain two spectral sections which can not be smoothly connected
 (Fig. \ref{abs}).\\
Therefore we fixed the X--ray absorption to
 N$_{\rm H}=2.5\times 10^{21}$ 
cm$^{-2}$ and varied the optical absorption.\\
Using A$_{\rm B} \sim  4.2$,  U--filter data lie on the X--ray extrapolation,
but the optical spectrum would be very hard and the connection discontinuous.  
Using an intermediate  value, A$_{\rm B}=2.5$,
 the optical and X--ray spectra connect continuously on a parabolic
curve, evidenced in Fig. \ref{abs}. 
With this absorption value, we obtain a hard optical spectrum
($\alpha_{Jul}=0.94\pm0.01$ and $\alpha_{Oct}=0.64\pm0.02$, 
for the summer and the autumn observations respectively)
 and consequently higher synchrotron peak frequencies. 
A  dust--to--gas ratio higher than that reported by Ryter (1996)
 could therefore account for the observed optical to X--ray misalignement.\\
We can compare this behaviour with other multiwavelength campaigns:
in Fig. \ref{SED} we plot all the published SEDs of BL Lacertae.
Only in the SED of 1980 and in that of June 1999 we observe 
 synchrotron emission both in the optical and in the X--ray band.
In all the other cases, only the Compton component was detected, thus
one can not see any misalignement between the optical 
and X--ray synchrotron components.
While the {\it Beppo}SAX synchrotron spectrum of June 1999
connect continuously with the simultaneous optical data
(using the galactic dust--to--gas ratio; Ravasio et al., 2002),
 in the 1980 SED the optical--UV data 
seem to lie below  the extrapolation of the Einstein IPC spectrum
 (Bregman et al., 1990), 
recreating the misalignement observed in October--November 2000.\\ 
During the 1980 IPC observation, in fact, BL Lac was displaying a
  very steep synchrotron spectrum in the [0.1--4] keV range
 ($\alpha=2.2\pm1.0$, using however a very high absorption parameter
N$_{\rm H} = 1.25\times10^{22}$ cm$^{-2}$; Bregman et al., 1990),
hardly connectable with the very soft simultaneous optical--UV spectrum. 
The Einstein data, however, are  affected by very large uncertainties, which
make difficult to determine the exact  shape of the SED: this is evidenced
by the results published by Worrall \& Wilkes (1990),
 which reported an IPC spectral index
 $\alpha=1.34^{+2.5}_{-1.3}$. The uncertainties on the IPC spectral index
are such that no firm conclusion about the reality of the
optical--to--X--ray misalignement in the 1980 SED can be drawn.\\
The strange optical/X--ray misalignement observed in  2000
 (and maybe in 1980) is not detected in the only other 
 multiwavelength campaign that shows synchrotron emission in the optical
and X--ray bands (June 1999, {\it Beppo}SAX).
 Nevertheless, we can rely on the goodness of our data:
 the {\it Beppo}SAX spectrum is confirmed 
by simultaneous RXTE data, as described in the
previous sections, while the optical data are confirmed
 by different observatories (Villata et al., 2002).
 The optical/X--ray misalignement is therefore real.
The absorption along the line of sight can not account for it, 
unless we assume a sudden, large increase of the dust--to--gas ratio
in the interstellar material.\\
This hypothesis, however, is not unlikely, since
the line of sight towards BL Lac is 
partially covered by a low surface brightness interstellar nebulosity
which is very variable (Penston \& Penston, 1973; Sillanp\"a\"a, priv. comm.).
If these  clouds are dusty, they could account for the 
dust--to--gas ratio excess needed to reconcile the optical and X--ray data 
in the SED of October--November 2000 (and possibly also in  the  1980
SED; Sillanp\"a\"a et al., 1993).
Their proper motion
could indeed explain the misalignement
in the SED of autumn 2000 and its absence in that of  June 1999.

\subsection{Bulk Compton radiation}
\begin{figure}
\begin{center}
\psfig{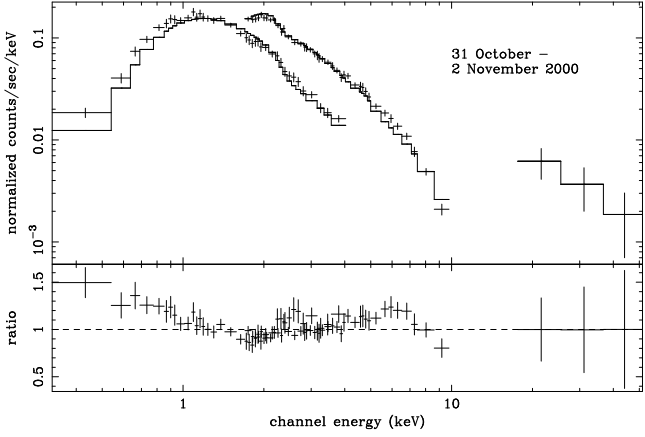}
\caption{LECS+MECS+PDS 31-October -- 2 November spectrum of BL Lac.
fitted by the model in equation \ref{newmod}, using $\Gamma_1=1$.
If $\Gamma_1 \leq 1$ the data are badly fitted by this model.}
\label{bbody}
\end{center}
\end{figure}

Alternatively, the observed misalignement could be claimed 
as the first detection of the so called ``bulk Compton'' emission.\\
This possibility  was predicted by Sikora et al. (1997): he postulated the
existence of a population of ``cold'' electrons in addition to the 
relativistic ones producing synchrotron and inverse Compton emission.
 These cold electrons, however,  have bulk relativistic motion
  with respect to the radiation fields produced by the accretion disk 
and by the broad line region and can 
inverse Compton scatter these photons up to frequencies
\begin{equation}
\nu_{\rm BC} \simeq \Gamma^2 \nu_0
\label{nubc}
\end{equation}
where $\nu_0$ is a characteristic frequency  of the external
 radiation field and $\Gamma$ is the bulk Lorentz factor of the jet.
If the seed radiation peaks in the UV band, as the accretion disk
 and the broad line region emission do, the  bulk Compton  component 
will peak in the soft X--ray band
 accounting for the misalignement we observe in the 2000 BL Lac SED.\\
Assuming a conical jet with opening angle $\theta_j\sim 1/\Gamma$
and the conservation of the flux of electrons along the jet 
($n_e\propto1/r^2$), Sikora et al.(1997) estimated the observed 
amount of bulk Compton radiation $L_{\rm BC}$:
\begin{equation}
L_{\rm BC} \sim 2\Gamma^2 ~\frac{n_e(r_{\rm min}) r_{\rm min} \sigma_T ~\zeta L_{\rm UV}}{4}
\end{equation}  
where $r_{\rm min}$ is the distance from the apex of the cone at which the jet
is fully developed, 
$L_{\rm UV}$ is the external radiation field and $\zeta=\zeta(r)$
 is the fraction of the UV radiation contributing to the isotropic field
 at a distance $r$ from the jet  apex (Sikora et al., 1997).\\
In October--November the X--ray  luminosity was
 $L_{\rm X}=10^{45}$ erg s$^{-1}$.  
Assuming that  the bulk Compton emission peaks at $\nu = 10^{17}$ Hz
 and $\nu_0 = 10^{15}$ Hz, from Eq. \ref{nubc} we have $\Gamma_j \sim 10$.\\
Since Vermeulen et al. (1995) observed broad H$\alpha$ and H$\beta$
emission lines (confirmed by Corbett et al., 2000),
from their data  we can evaluate $L_{\rm UV}$ in an indirect way.
 Postulating a fixed line ratio
and following the method described in  Celotti, Padovani \& Ghisellini (1997)
 we obtain an average value L$_{\rm BLR}\sim5\times10^{42}$ erg s$^{-1}$.
If the  Broad Line Region covering factor is $\sim 10\%$, 
the disk luminosity is L$_{\rm UV}\sim 5\times 10^{43}$ erg s$^{-1}$.\\
Using these values and assuming $L_{\rm BC}=L_{\rm X}$ we obtain
the particle number density in the observer frame
\begin{equation}
n_e(r_{\rm min}) \simeq \frac{6\times10^9}{\Gamma^2_{10} \zeta_{0.1} ~r_{\rm min,15}} ~{\rm cm}^{-3}
\end{equation}
where $\Gamma_{10} = \Gamma$/10, $\zeta_{0.1}=\zeta/0.1$
and $r_{\rm min,15}=r_{\rm min}/10^{15} {\rm cm}$.\\
This number particle density,  necessary to 
produce the observed X--ray spectrum,  puts constraints
 on the jet composition.
In fact, if we suppose a one--to--one proton--electron plasma,
  we can easily evaluate the total kinetic power of the jet
(Celotti \& Fabian, 1993):
\begin{eqnarray}
 L_{\rm KIN} \sim \pi (r_{\rm min} \theta_j)^2 n'_p m_p c^2 \Gamma^2 c \sim 
 \nonumber\\ 
\sim 10^{47} ~\frac{r_{\rm min,15} ~\theta^2_{j,0.1}}{\Gamma_{10} ~\zeta_{0.1}}
 ~{\rm erg ~s}^{-1}
\end{eqnarray}
where $n'_p = n_p/\Gamma$ is the proton number density 
in the jet comoving frame. This value exceeds the $L_{\rm KIN}$ estimated
by Celotti, Padovani \& Ghisellini (1997) by a factor $\sim 30$.
To produce the observed X--ray spectrum via bulk Compton mechanism
and to be consistent with the result of Celotti, Padovani \& Ghisellini (1997)
the proton--electron ratio must be smaller: the jet must be pair loaded.\\
This is not forbidden since the  optical thickness
 for Thomson scattering ($\tau(r) = n_e \sigma_T \theta_j r$) 
at $r_{\rm min}$ is 
\begin{equation}
\tau(r_{\rm min})=\frac{2}{5} \frac{\theta_{j,0.1}}{\Gamma^2_{10} ~\zeta_{0.1}} 
\end{equation}
Since from our calculations $\tau(r_{\rm min}) < 1$ 
(and  decreasing  further out)
and the expansion timescale at $r_{\rm min}$
  ($t'_{\rm exp}\sim 8\times10^4 r_{\rm min, 15} \Gamma_{10}$ s)
is smaller than the annihilation timescale
 ($t_{ann} \sim 2\times 10^5 ~\Gamma^3_{10} ~\zeta_{0.1} r_{\rm min,15}$ s)
the  pairs will survive along the jet.\\
After having ruled out the hypothesis of a proton--electron jet 
and having requested the presence of pairs in the jet,
it is now fundamental to understand if the particle number necessary
to produce the bulk Compton emission is sufficient to produce
the observed SEDs. 
The particle number  conservation we have just demonstrated implies 
that $n_e \propto 1/r^2$. 
Therefore, at the distance where the radiation is emitted, 
$r_{\rm SED} \sim 10^{16} - 10^{17}$ cm, the particle number density would be
\begin{eqnarray}
n(r_{\rm SED}) = n(r_{\rm min}) \Big( \frac{r_{\rm min}}{r_{\rm SED}} \Big)^2 \sim 
 \nonumber\\
\sim 6\times 10^7 \frac{1}{\Gamma^2_{10} ~\zeta_{0.1}}\frac{r_{\rm min,15}}{r^2_{\rm SED,17}}  ~{\rm cm}^{-3}
\end{eqnarray}
This value is greater than what we needed 
to model old BL Lac seds: $n_e \sim 10^{7}/\gamma_{\rm min}$ cm$^{-3}$.
Only a fraction of the electrons carried by the jet need to be accelerated
in order to produce the observed radio to $\gamma$--ray emission.\\  
What we have proved to this point is that the bulk Compton can 
account for the X--ray data as long as the jet is pair rich. 
Further constraints about this emission mechanism  come from 
the investigation of the pair loading processes occurring in the protojet,
e.g. photon--photon interaction (Svensson, 1987) or
the interaction with the X--ray corona field (Sikora \& Madejski, 2002),
 but this is beyond the goal of this paper.\\
However a simple viability test of the bulk Compton scenario can be performed 
which is based on the X--ray spectral shape.
If this mechanism is working, in the X--ray band we should see 
the exponential tail of the  blueshifted multi--temperature blackbody emission
 of the accretion disk superposed to the usual hard inverse Compton emission.
Therefore we tried to fit {\it Beppo}SAX data with a new model
described by:
\begin{equation}
A(\nu) = e^{-\rm N_{H} \sigma(\nu)}(k_1\nu^{-\Gamma_1} e^{-[(\nu_0-\nu)/\nu_1]}+k_2\nu^{-\Gamma_2})
\label{newmod}
\end{equation}
where the second power law parameters are 
 the best fit values obtained from modelling PDS
 data alone with a power law model
 ($\Gamma_2 = 1.76$; $k_2 =3.19\times 10^{-3}$  photons/keV/cm$^2$/s); 
N$_{\rm H}$ was fixed to $2.5\times 10^{21}$ cm$^{-2}$.
This model is not able to reproduce the data unless $\Gamma_1 >1$ 
(see e.g. in fig. \ref{bbody} the residuals below 1 keV).
 This implies that a blueshifted 
Shakura--Sunyaev disk spectrum cannot  reproduce 
the observed soft X--ray spectrum
unless being characterized by a soft power law slope in the UV band.\\
\subsection{Two synchrotron components}
We considered then the hypothesis that the optical and the X--ray spectra
we observed are created by  two populations of emitting electrons
possibly  located in two regions at different
distances from the nucleus.
One  produces a synchrotron component peaking in the IR band and
 is responsible for the WEBT section 
while the second accounts for the {\it Beppo}SAX data.
In order not to overproduce the optical flux, the X--ray component
 must peak above the UV band. This implies that the 
emitting particle population must have a break ($\gamma_{\rm break}$).
If $N(\gamma) \propto \gamma^{-2}$ below $\gamma_{\rm break}$,
we require 
$\gamma_{\rm break} > 1.6 \times 10^{4} ~\delta^{-1/2}_{10} B^{-1/2}$
 to not overproduce the optical flux.\\
The observed  fast X--ray variability implies that the 
presence of a break in the particle distribution could not be 
an effect of the  cooling.
Since $t_{\rm var}\sim 6\times10^3$ s (see table \ref{var})
the X--ray emitting region must be located at  a distance from the nucleus 
$R\lesssim \delta ~c ~t_{\rm var}/ \psi \sim 2\times 10^{16} ~\Gamma^2_{10}$
 cm, where $\psi \sim 1/\Gamma$ is the jet opening angle  and 
$\delta = 1/(\Gamma (1-\beta \cos{\theta_{obs}}))$ is the Doppler factor.
The dimension of the BLR of BL Lac is 
$R_{\rm BLR} \sim 8.3 \times 10^{16} ~{\rm M}_{BH,8}$ cm
 (B\"ottcher \& Bloom, 2000),
where ${\rm M}_{BH,8}$ is the black hole mass in units of $10^8$ solar masses.
Since M$_{BH,8} \sim 1.7$ (Woo \& Urry, 2002),
 $R_{\rm BLR} \sim 1.4 \times 10^{17}$ cm:
the X--ray emitting  region should be inside the Broad Line Region, 
where the cooling rate is high.\\
If we assume  that  a continuous 
power law distribution of particles N$(\gamma) \propto \gamma^{-n}$
is injected for a time $t_{\rm var}$, then  an electron above $\gamma_c$  
can cool in a time  $t_{\rm var}$. 
After this time, above $\gamma_c$ the particle population
steepens to N$(\gamma) \propto \gamma^{-(n+1)}$, 
while below $\gamma_c$ it remains unchanged.
The value of  $\gamma_c$ is:
\begin{eqnarray}
\lefteqn{\gamma_{\rm c} \sim \frac{3 m_e c^2}{4 \sigma_T c ~U'_{\rm BLR}(1+U'_B/U'_{\rm BLR}+U'_{\rm s}/U'_{\rm BLR})~\delta t_{\rm var}} \sim}
\nonumber\\ 
 & & \sim 7 \times 10^3 \frac{1}{\Gamma^2_{10} ~\delta_{10}}
\end{eqnarray} 
where we have used $U'_{BLR}= 7\times10^{-2} ~\Gamma^2_{10}$ erg cm$^{-3}$.
We can see that $\gamma_c$ is hardly consistent with 
 the required $\gamma_{\rm break}$, but, if during our observation
the luminosity of the lines was fainter than the average L$_{\rm BLR}$
we have considered, then it is possible that 
$\gamma_c \approx \gamma_{\rm break}$. In this case
the required break in the distribution 
can be accounted by the radiative cooling.
Otherwise,  not to overproduce the optical flux,
the X--ray emitting particle distribution should be injected
 with an intrinsic break.
%
%
%
%
%
%
%
%
\subsection{Klein-Nishina effect on the synchrotron spectrum}
Since the Compton cooling rate of electrons with energies
\begin{equation}
\gamma > \gamma_1 = {m_e c^2 \over \Gamma h\nu_{BEL}} \simeq 
{10^4 \over \Gamma_{10} \nu_{BEL,15}} 
\end{equation}
is reduced because of the decline of the Compton scattering cross section
in the Klein--Nishina regime, for a given electron injection rate, 
the energy distribution of electrons hardens beyond $\gamma_1$.
The effect is particularly strong for $f \equiv U_{BLR}'/U_B' \gg 1$ and 
causes flattening of the synchrotron spectrum at 
\begin{eqnarray}
\lefteqn{ \nu > \nu_1 \simeq 3.6 \times 10^6 ~{\rm Hz} \, B ~\gamma_1^2 ~\delta 
 \simeq } \nonumber\\
 &  &  \simeq  5 \times 10^{15}  { (U_{BLR}/0.005~{\rm erg\,cm}^{-3})^{1/2} 
(\delta/\Gamma)
\over (f/10)^{1/2} \nu_{BEL,15}^2}  ~{\rm Hz}
\end{eqnarray}
The hardened part of the electron energy distribution extends up to
$\gamma_2 \sim f^{1/2} \gamma_1$. At $\gamma > \gamma_2$ the
cooling of electrons
becomes to be dominated by synchrotron radiation and the slope of
the electron energy distribution becomes the same as at $\gamma \ll \gamma_1$.
Consequently, at $\nu > \nu_2 \sim f \nu_1$, the  
synchrotron spectrum steepens, regaining the slope from the optical band but 
vertically shifted up by a factor $f$ above the extrapolation of 
the optical spectral portion. 
Such a scenario can reproduce  the observed spectral 
``glitch'', provided $f \sim 10$. Details of the model will be presented 
elsewhere; here we would like to mention only that a similar scenario has 
been proposed by
Dermer and Atoyan (2002) to explain spectral glitches between the optical and 
X-ray spectra of kiloparsec scale jets.
\section{Conclusions}
BL Lac has been the target of a multiwavelength campaign during 2000,
extending from June to November. {\it Beppo}SAX observed the source 
twice, in July and at the end of October, while in different optical state
of activity. As evidenced also from previous observations,
BL Lac displays a very complex behaviour.
We summarize the {\it Beppo}SAX results as follows:
\begin{itemize}
\item In July, while optically weak, the source displayed a faint, 
hard X--ray spectrum
with  positive residuals towards low energies.
The soft X--ray flux varied in a timescale of a few hours,
 while the hard X--ray flux was almost constant.
\item In October, while BL Lac was bright in the optical band,
  we observed in the X--ray the transition from
an extremely  strong, soft component to a hard 
component dominating above 10--20 keV. The soft spectrum displayed
an erratic temporal behaviour with large and fast variations on timescales 
down to $6\times 10^3$ s, while the hard component remained almost constant.
\item {\it Beppo}SAX spectral results are confirmed by simultaneous
RXTE data: PCA and MECS spectra are consistent within the uncertainties.
\end{itemize}
The frequency dependent temporal variability is consistent with the 
spectral analysis results: we observed fast and large flux variations
when the soft component, interpreted as synchrotron emission, was dominating,
 while we observed constant light curves in the hard section of the spectra
which can be reproduced as inverse Compton radiation in a leptonic jet model.
 This can be explained
since synchrotron X--ray emitting electrons are more energetic than
those that are producing inverse Compton emission at X--ray energies,
 so they cool faster.
This behaviour was already observed in BL Lac (Ravasio et al., 2002) 
as well as in other similar objects, such as ON 231 (Tagliaferri et al., 2000)
or S5~ 0716+714 (Tagliaferri et al., 2003).\\
The analysis of the multiwavelength Spectral Energy Distributions and
the comparison with other historical SEDs,  evidences the exceptionality
 of the X--ray spectrum of October 2000: during this observation
BL Lac was displaying the highest soft X--ray flux ever recorded
and an integrated [2-10] keV flux which was only sligthly smaller 
than that detected in July 1997, while BL Lac was in an exceptional 
flaring state and was displaying a hard X--ray spectrum 
(Madejski et al., 1999).\\
Moreover, the SED of October 2000 displayed another very interesting feature:
the soft X--ray data laid above the extrapolation of the optical spectrum,
while they should be both produced via the same synchrotron emission.
To account for this inconsistency we have investigated 4 possibilities,
among which, however, we cannot at present discriminate:
\begin{enumerate}
\item The dust--to--gas ratio towards BL Lac is higher than the 
interstellar one.  However, since this misalignement
is not seen in the other multiwavelength SED of BL Lac
with a well defined  synchrotron component
 (except that of 1980,
where data have large uncertainties),  a sudden large increase
of the dust--to--gas ratio is  needed. This hypothesis is not unlikely
since  dusty galactic nebulosities are indeed  observed towards 
the source.

\item This is the first detection of the so called bulk Compton 
emission. In the  hypothesis  that the jet is pair loaded,
we can reproduce the observed soft X--ray spectrum  via the 
bulk Compton mechanism
while the optical and the hard X--ray spectra can be 
explained via the usual synchrotron and inverse Compton models.
However, a blueshifted  Shakura--Sunyaev disk spectrum
cannot model {\it Beppo}SAX data. Nevertheless, we cannot exclude this scenario
since the effective spectral shape of accretion disks in the far UV
 is still uncertain
and  could have a power law rather than an exponential slope
(see e.g. Laor et al., 1996).
\item Two synchrotron components present at different distances from the 
nucleus account for the optical and the X--ray spectra, respectively.
In order to not overproduce the optical flux and to account for the X--ray 
fast variability, the X--ray emitting particle population
 must have a break at energies greater than 
$\gamma \sim 1.6 \times 10^4 \delta^{-1/2}_{10} B^{-1/2}$.
The break can be accounted by the particle cooling
 or must be intrinsic if the line luminosity level 
is lower or higher than the calculated average value, respectively. 
\item Through the synchrotron emitting population,
 we  observe the transition from inverse Compton cooling dominated
to synchrotron cooling dominated particles.
 Between these two conditions, there is a range in which the  electrons
 are dominated by an inefficient inverse Compton cooling 
in the Klein--Nishina regime: this produces a hardening in the particle
population and therefore in the synchrotron spectrum which
accounts for the observed optical to X--ray misalignement.
\end{enumerate}
 In the SED of October 2000 of BL Lac, above the synchrotron peak, 
 we have detected a well determined misalignement between 
two sections of the synchrotron spectrum. 
To explain this feature, we suggested four scenarios
 presenting different favorable and critical points:
at present, however, the theoretical models and the data do not allow us to 
discriminate between them, although the hypothesis of a variable 
dust--to--gas ratio is the most plausible one.
\\
\begin{acknowledgements}
We are grateful to Dr. M. Villata for sending us informations
about the optical data published in Villata et al. (2002).
This research was financially supported by the 
Italian Ministry for University and Research.
M.S. acknowledges partial support from Polish KBN grants:
5P03D00221 and 2P03C006 19p1,2.
\end{acknowledgements}

\end{document}